\documentclass[preprint,10pt]{aastex}
\usepackage{pdflscape}
\usepackage{geometry}
\usepackage{tablefootnote}

\begin{document}

\title{Testing the Asteroseismic Mass Scale Using Metal-Poor Stars Characterized with APOGEE and \textit{Kepler}}
\shorttitle{Testing the Asteroseismic Mass Scale}
\author{
Courtney R.\ Epstein\altaffilmark{1},
Yvonne P.\ Elsworth\altaffilmark{2},
Jennifer A.\ Johnson\altaffilmark{1,3}
Matthew Shetrone\altaffilmark{4},
Beno\^{i}t Mosser\altaffilmark{5},
Saskia Hekker\altaffilmark{6},
Jamie Tayar\altaffilmark{1},
Paul Harding\altaffilmark{7},
Marc Pinsonneault\altaffilmark{1},
V\'{i}ctor Silva Aguirre\altaffilmark{8},
Sarbani Basu\altaffilmark{9},
Timothy C.\ Beers\altaffilmark{10,11},
Dmitry Bizyaev\altaffilmark{12},
Timothy R.\ Bedding\altaffilmark{13},
William J. Chaplin\altaffilmark{2},
Peter M. Frinchaboy\altaffilmark{14},
﻿Rafael A.\ Garc\'{i}a\altaffilmark{15},
Ana E.\ Garc\'{i}a P\'{e}rez\altaffilmark{16},
Fred R.\ Hearty\altaffilmark{16},
Daniel Huber\altaffilmark{17},
Inese I.\ Ivans\altaffilmark{18},
Steven R.\ Majewski\altaffilmark{16},
Savita Mathur\altaffilmark{19},
David Nidever\altaffilmark{20},
Aldo Serenelli\altaffilmark{21},
Ricardo P.\ Schiavon\altaffilmark{22},
Donald P. Schneider\altaffilmark{23,24},
Ralph Sch\"{o}nrich\altaffilmark{1,25},
Jennifer S.\ Sobeck\altaffilmark{11,26},
Keivan G.\ Stassun\altaffilmark{27},
Dennis Stello\altaffilmark{28},
Gail Zasowski\altaffilmark{1,2,29}
}
\altaffiltext{1}{Department of Astronomy, Ohio State University,
140 W.\ 18th Ave., Columbus, OH 43210, USA;
epstein@astronomy.ohio-state.edu}
\altaffiltext{2}{School of Physics \& Astronomy, University of Birmingham, Edgbaston Park Road, West Midlands, Birmingham, UK, B15 2TT}
\altaffiltext{3}{Center for Cosmology and Astro-Particle Physics, Ohio State University, Columbus, OH 43210, USA}
\altaffiltext{4}{McDonald Observatory, The University of Texas at Austin, 1 University Station, C1400, Austin, TX 78712-0259, USA}
\altaffiltext{5}{LESIA, CNRS, Universit\'e Pierre et Marie Curie, Universit\'e Denis Diderot, Observatoire de Paris, 92195 Meudon Cedex, France}
\altaffiltext{6}{Max-Planck-Institut f\"ur Sonnensystemforschung, Justus-von-Liebig-Weg 3, 37077 G\"ottingen, Germany}
\altaffiltext{7}{Department of Astronomy, Case Western Reserve University, Cleveland, OH 44106-7215, USA}
\altaffiltext{8}{Stellar Astrophysics Centre, Department of Physics and Astronomy, Aarhus University, Ny Munkegade 120, DK-8000 Aarhus C, Denmark}
\altaffiltext{9}{Department of Astronomy, Yale University, P.O. Box 208101, New Haven, CT 06520-8101, USA}
\altaffiltext{10}{National Optical Astronomy Observatory, Tucson, AZ 85719, USA}
\altaffiltext{11}{JINA:  Joint Institute for Nuclear Astrophysics}
\altaffiltext{12}{Apache Point Observatory, Sunspot, NM 88349, USA}
\altaffiltext{13}{Sydney Institute for Astronomy (SIfA), School of Physics, University of Sydney, NSW 2006, Australia}
\altaffiltext{14}{Department of Physics \& Astronomy, Texas Christian University, TCU Box 298840, Fort Worth, TX 76129}
\altaffiltext{15}{Laboratoire AIM, CEA/DSM-CNRS, Université Paris 7 Diderot, IRFU/SAp, Centre de Saclay, 91191, Gif-sur-Yvette, France }
\altaffiltext{16}{Department of Astronomy, University of Virginia, Charlottesville, VA 22904, USA}
\altaffiltext{17}{NASA Ames Research Center, Moffett Field, CA 94035, USA }

\altaffiltext{18}{Department of Physics and Astronomy, The University of Utah, Salt Lake City, UT 84112, USA}
\altaffiltext{19}{Space Science Institute, 4750 Walnut street Suite \#205, Boulder, CO 80301, USA}
\altaffiltext{20}{Department of Astronomy, University of Michigan, Ann Arbor, MI, 48109, USA}
\altaffiltext{21}{Institute of Space Sciences (IEEC-CSIC), Campus UAB, Fac. Ciències, Torre C5 parell 2, E-08193 Bellaterra, Spain }
\altaffiltext{22}{Astrophysics Research Institute, IC2, Liverpool Science Park, Liverpool John Moores University, 146 Brownlow Hill, Liverpool, L3 5RF, UK}
\altaffiltext{23}{Department of Astronomy and Astrophysics, The Pennsylvania State University, University Park, PA 16802}
\altaffiltext{24}{Institute for Gravitation and the Cosmos, The Pennsylvania State University, University Park, PA 16802}
\altaffiltext{25}{Rudolf-Peierls Centre for Theoretical Physics, University of Oxford, 1 Keble Road, OX1 3NP, Oxford, United Kingdom}
\altaffiltext{26}{Department of Astronomy \& Astrophysics, University of Chicago, 5640 S. Ellis Avenue, Chicago, Illinois, 60637}
\altaffiltext{27}{Department of Physics \& Astronomy, Vanderbilt University, Nashville, TN 37235, USA}
\altaffiltext{28}{Sydney Institute for Astronomy (SIfA), School of Physics, University of Sydney, NSW 2006, Australia}
\altaffiltext{29}{Department of Physics and Astronomy, Johns Hopkins University, Baltimore, MD}

\begin{abstract}

Fundamental stellar properties, such as mass, radius, and age, can be inferred using asteroseismology. Cool stars with convective envelopes have turbulent motions that can stochastically drive and damp pulsations.  The properties of the oscillation frequency power spectrum can be tied to mass and radius through solar-scaled asteroseismic relations. Stellar properties derived using these scaling relations need verification over a range of metallicities. Because the age and mass of halo stars are well-constrained by astrophysical priors, they provide an independent, empirical check on asteroseismic mass estimates in the low-metallicity regime. We identify nine metal-poor red giants (including six stars that are kinematically associated with the halo) from a sample observed by both the \textit{Kepler} space telescope and the Sloan Digital Sky Survey-III APOGEE spectroscopic survey. We compare masses inferred using asteroseismology to those expected for halo and thick-disk stars. Although our sample is small, standard scaling relations, combined with asteroseismic parameters from the APOKASC Catalog, produce masses that are systematically higher ($\left<\Delta \mathrm{M}\right>=0.17\pm0.05$ M$_\odot$) than astrophysical expectations.
The magnitude of the mass discrepancy is reduced by known theoretical corrections to the measured large frequency separation scaling relationship. Using alternative methods for measuring asteroseismic parameters induces systematic shifts at the 0.04 M$_\odot$ level.
We also compare published asteroseismic analyses with scaling relationship masses to examine the impact of using the frequency of maximum power as a constraint. Upcoming APOKASC observations will provide a larger sample of $\sim100$ metal-poor stars, important for detailed asteroseismic characterization of Galactic stellar populations.
 \end{abstract}
\keywords{asteroseismology---Galaxy: halo---stars: fundamental parameters}

\section{Introduction}

Accurate determinations of fundamental stellar properties are required to improve our understanding of stellar populations and Galactic formation. Inferring these properties is notoriously difficult, unless stars are members of clusters or eclipsing binary systems. 
However, we can probe stellar interiors through global oscillations.
After several ground-based studies \citep{Bedding2011_Observational_Perspective} and serendipitous space-based observations (HST and WIRE; e.g.\ \citealt{Stello2009} and references therein), the space-based telescopes CoRoT \citep{Michel2008} and \textit{Kepler} \citep{Borucki2010} made asteroseismic characterization possible for thousands of stars. In an asteroseismic analysis, the average spacing between consecutive overtones of the same angular degree (average large frequency separation, $\Delta\nu$) and the peak in the Gaussian-like envelope of mode amplitudes (frequency of maximum oscillation power, $\nu_\mathrm{max}$) is derived from the frequency power spectrum. For oscillations driven by surface convection, empirical scaling relations \citep[hereafter SRs;][and references therein]{Kjeldsen1995} connect these asteroseismic observables to mass, radius, and effective temperature:
\begin{eqnarray}
  \frac{\Delta \nu}{\Delta \nu_{\odot}} \simeq& \left(\frac{\mathrm{M}}{\mathrm{M}_{\odot}}\right)^{1/2} \left(\frac{\mathrm{R}}{\mathrm{R}_{\odot}}\right)^{-3/2}\label{eq:delta nu scaling relation} \\
  \frac{\nu_\mathrm{max}}{\nu_{\mathrm{max},\odot}} \simeq& \left(\frac{\mathrm{M}}{\mathrm{M}_{\odot}}\right)\left(\frac{\mathrm{R}}{\mathrm{R}_\odot}\right)^{-2}\left(\frac{\mathrm{T}_\mathrm{eff}}{\mathrm{T}_{\mathrm{eff,\odot}}}\right)^{-1/2}, \label{eq:nu max scaling relation}
\end{eqnarray}
where $\Delta \nu_\odot=135.0 \pm 0.1$ $\mu$Hz, $\nu_{\mathrm{max},\odot}=3140 \pm 30$ $\mu$Hz, and T$_{\mathrm{eff},\odot}=5777$ K (Pinsonneault et al., \textit{in prep.}). Solving for mass and radius yields
\begin{eqnarray}
  \frac{\mathrm{M}}{\mathrm{M}_{\odot}}\simeq& \left(\frac{\nu_\mathrm{max}}{\nu_{\mathrm{max},\odot}}\right)^3 \left(\frac{\Delta \nu}{\Delta \nu_{\odot}}\right)^{-4} \left(\frac{\mathrm{T}_\mathrm{eff}}{\mathrm{T}_{\mathrm{eff},\odot}}\right)^{3/2} \label{eq:mass scaling relation} \\
  \frac{\mathrm{R}}{\mathrm{R}_{\odot}}\simeq& \left(\frac{\nu_\mathrm{max}}{\nu_{\mathrm{max},\odot}}\right) \left(\frac{\Delta \nu}{\Delta \nu_{\odot}}\right)^{-2} \left(\frac{\mathrm{T}_\mathrm{eff}}{\mathrm{T}_{\mathrm{eff},\odot}}\right)^{1/2}. \label{eq:radius scaling relation}
\end{eqnarray}
The SRs take no account of metallicity dependence and they were developed for stars like the Sun, so it is not obvious that they should work for red giant branch (RGB) stars, which have a different internal structure. There are observational and theoretical problems with defining and measuring $\nu_\mathrm{max}$ and $\Delta\nu$.

Empirical tests of the radius and mass from SRs have been restricted to metallicities near solar ($-0.5\lesssim\mathrm{[Fe/H]}\lesssim +0.4$). Asteroseismic radii agree within $<$5\% when compared with interferometry \citep{Huber2012}, \textit{Hipparcos} parallaxes \citep{SilvaAguirre2012}, and RGB stars in the open cluster NGC6791 \citep{Miglio2012}.
SR masses are less precise than SR radii and fundamental mass calibration is also intrinsically more difficult.
\citet{Brogaard2012} anchored the mass scale of the super-solar cluster NGC6791 to measurements of eclipsing binaries at the main-sequence turn-off (MSTO) and inferred M$_\mathrm{RGB} = 1.15 \pm 0.02$ M$_\odot$, lower than masses derived from standard SRs (M$_\mathrm{RGB}=1.20 \pm 0.01$ M$_\odot$ and $1.23 \pm 0.02$ M$_\odot$ from \citealt{Basu2011} and \citealt{Miglio2012}, respectively). This is not conclusive evidence that the SR are in error because the mass estimates are sensitive to temperature scale and bolometric corrections. Even using a new less-temperature sensitive SR, \citet{Wu2014} found M$_\mathrm{RGB}=1.24\pm0.03$ M$_\odot$ in NGC6791.

The \textit{Kepler} Asteroseismic Science Consortium (KASC) detected solar-like oscillations in 13,000+ red giants \citep[e.g.,][]{Stello2013}. As part of the Sloan Digital Sky Survey III \citep[SDSS-III;][]{Eisenstein2011}, the Apache Point Observatory Galaxy Evolution Experiment (APOGEE; Majewski et al., \textit{in prep.}) is obtaining follow-up spectra of these asteroseismic targets. APOGEE uses a high-resolution ($R\sim 22,500$), $H$-band, multi-object spectrograph whose seven square-degree
field-of-view \citep{Gunn2006} is well-matched to the size of one of \textit{Kepler}'s 21 CCD modules. The APOKASC Catalog (Pinsonneault et al., \textit{in prep.}) reports asteroseismic and spectroscopic results for stars in the \textit{Kepler} field observed in APOGEE's first year of operations.

Pinsonneault et al.\ (\textit{in prep.})
describe the asteroseismic analysis, including the preparation of raw \textit{Kepler} light curves
\citep{Garcia2011}, measurement of $\Delta \nu$ and $\nu_\mathrm{max}$, and outlier rejection procedures. 
We used up to five methods to extract $\Delta\nu$ and $\nu_\mathrm{max}$ from the frequency-power spectrum (\citealt{Huber2009,Hekker2010}, OCT; \citealt{Kallinger2010,Mathur2010,Mosser2011}, COR). 
Because the OCT method had the highest overall completion fraction, the APOKASC Catalog reports $\Delta \nu$ and $\nu_\mathrm{max}$ from OCT with uncertainties that combine, in quadrature, the formal OCT uncertainty, the standard deviation of results from all methods, and an allowance for known issues with the SR \citep[e.g.][]{Miglio2012}.

We perform with the APOKASC sample the first test of asteroseismic SR mass estimates in the low-metallicity regime where strong priors on stellar ages and masses exist. For this, we identify rare halo stars, explicitly targeting high--proper motion stars and low-metallicity candidates selected using Washington photometry (Harding et al., \textit{in prep.}) and low-resolution spectroscopy.

\section{The Metal-Poor Sample}\label{sec:Halo Sample}

We identified nine stars among the 1,900 red giants in the APOKASC Catalog with [M/H$]<-1$, measured asteroseismic parameters, and no `BAD' spectroscopic parameters (Table \ref{table:Results}). Spectroscopic properties in the Catalog were derived using the APOGEE Stellar Parameters and Chemical Abundances Pipeline (ASPCAP; Garc{\'{\i}}a P{\'e}rez et al.\ \textit{in prep.}). ASPCAP used $\chi^2$ minimization in a library of synthetic spectra to find the combination of temperature (T$_\mathrm{eff}$), surface gravity ($\log g$), metallicity ([M/H]$_\mathrm{ASPCAP}$), carbon ([C/M]), nitrogen ([N/M]), and alpha-element abundance ([$\alpha$/M]) that best reproduced the observed spectrum. We adopt the calibration from \citet{Meszaros2013}, who 
compared the raw ASPCAP stellar parameters with well-studied clusters and asteroseismic $\log g$ and corrected raw ASPCAP metallicities to reflect [Fe/H]. We also performed a line-by-line analysis of Fe, C, O, Mg, Si, and Al for the three most metal-poor stars, confirming the ASPCAP stellar parameters and abundances.

\begin{figure}
  \plottwo{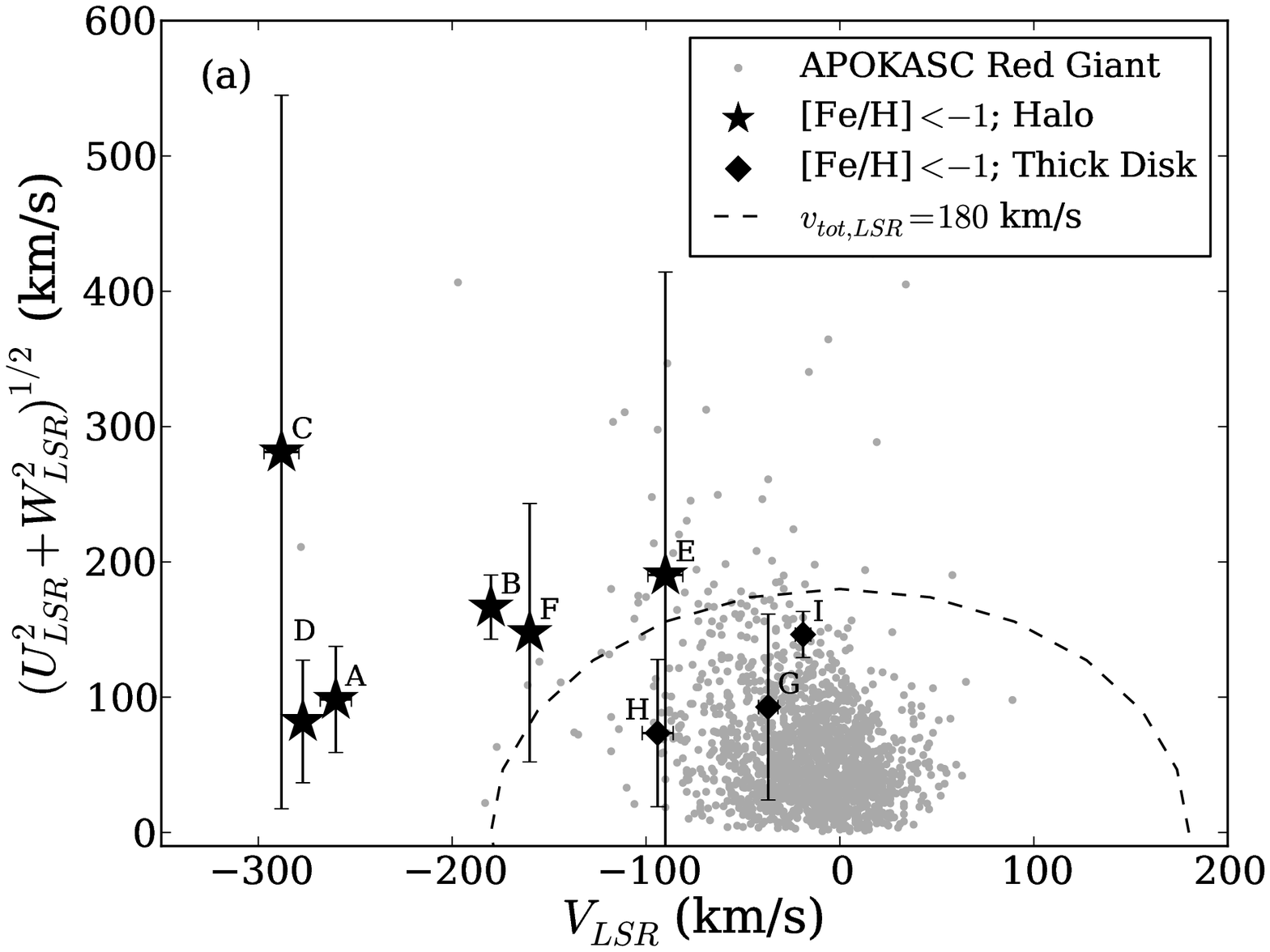}{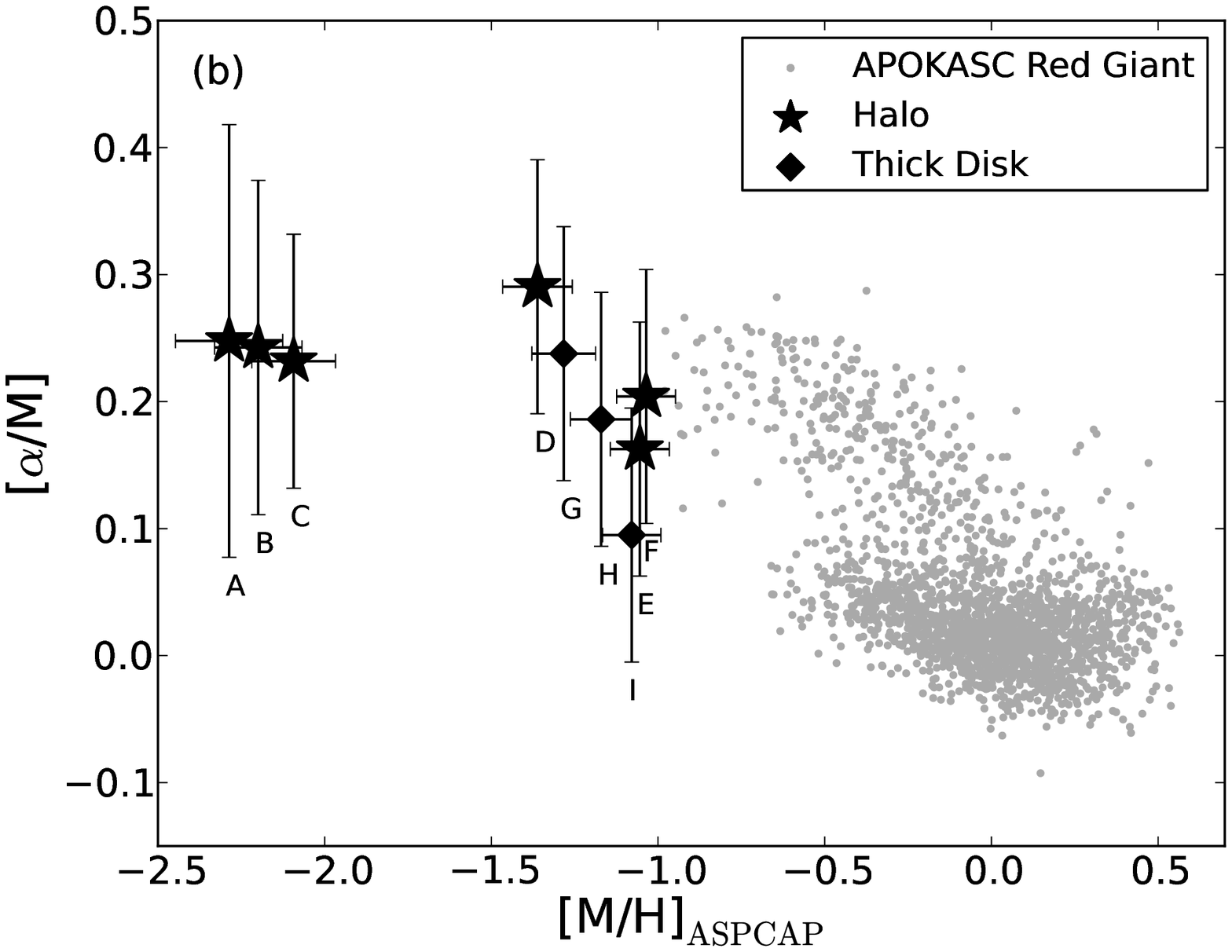}
  \caption{\textit{Left:} Toomre diagram of red giants with seismic detections in the APOKASC sample, using $v_{tot,LSR}=180$ km/s as the kinematical division between stars classified as halo and disk \citep{Venn2004}. \textit{Right:} Metal-poor stars have enhanced ASPCAP [$\alpha$/Fe].}\label{fig:alpha}
\end{figure}

We used kinematics to discriminate between halo and disk populations. APOGEE measures radial velocities accurate to better than 150 m/s (Nidever et al., \textit{in prep}), and UCAC-4 \citep{Zacharias2013}
reports proper motions. Adopting \citet{Schlegel1998} extinctions and \citet{Bressan2012} bolometric corrections, we calculated luminosities from SR radii and ASPCAP T$_\mathrm{eff}$s and then distances from the 2MASS $J$-band magnitude \citep{Skrutskie2006}. Three-dimensional space velocities $(U, V, W)$ were derived using the \citet{JohnsonSoderblom1987} prescription, correcting for the solar motion relative to the Local Standard of Rest \citep[LSR;][]{Schonrich2010}.
From a Toomre diagram (Figure \ref{fig:alpha}a), the four most metal-poor stars in our sample show halo-like kinematics, KIC11181828 and KIC5858947 are ambiguous, and the remaining three show thick-disk-like kinematics.

\section{Establishing Age and Mass Expectations}\label{sec:Expectations and Models}

Independent measurements of the mass and age of halo and thick-disk stars determined the astrophysical priors that we adopted for the mass of metal-poor APOKASC stars. The best direct mass constraint for a metal-poor star comes from a MSTO eclipsing binary in the thick-disk globular cluster (GC) 47 Tuc, with measured masses M$_p=0.8762\pm0.0048$ M$_\odot$ and M$_s=0.8588 \pm 0.0060$ M$_\odot$ \citep{Thompson2010}.

Age constraints are more readily available than direct mass determinations, and provide an indirect constraint on halo-star masses. GC ages are inferred by determining the color and luminosity of the MSTO \citep[e.g.,][]{Gratton1997}, or by using the white dwarf cooling sequence \citep[e.g.,][]{Hansen2002}. Field-star ages have been derived using the masses of local white dwarfs \citep{Kalirai2012}, and the imprint left by the MSTO on the stellar temperature distribution function \citep{Jofre2011}. All of these methods indicate that the Galactic halo is 10 Gyr or older, with a mean value around 12 Gyr.

All nine of our metal-poor stars are $\alpha$-rich (Figure \ref{fig:alpha}b), in agreement with the range of [$\alpha$/Fe] seen for halo GCs \citep{Meszaros2013}. Thick-disk stars are as $\alpha$-enhanced as the halo stars, suggesting that they also formed before SNIa significantly polluted the interstellar medium, which occurred 1-1.5 Gyr after star formation began \citep[e.g.,][]{Matteucci2009}. From a volume-limited sample of \textit{Hipparcos} stars, \citet[][and references therein]{Fuhrmann2011} argued that the thick disk formed in a single-burst 12-13 Gyr ago, followed by a gap in star formation. \citet{Haywood2013} found some $\alpha$-enhanced, metal-poor disk stars as old as 13 Gyr, but none younger than 8 Gyr. The age of the Universe \citep[$13.77\pm0.059$ Gyr;][]{Bennett2012} provides the upper bound on age.
 
Translating age constraints into mass expectations for metal-poor stars requires stellar models, which depend on the adopted helium and heavy-element mixture. We expect these nine stars to have a normal (near-primordial) helium abundance because our line-by-line spectroscopic analysis did not reveal Al-enhancements such as those seen in He-rich, second-generation GC stars \citep[e.g.,][]{Ventura2008}. To establish mass expectations, we consider only first ascent RGB models; in \S\ref{sec:Stellar Populations}, we discuss more evolved stars. We defined the range of expected halo-star masses and uncertainties by adopting $\pm2\sigma$ ranges of $+0.2$ to $+0.4$ dex for $[\alpha/\mathrm{Fe}]$, 10 to 13.77 Gyr for the age of the halo, and 8 to 13.77 Gyr for the age of the thick disk. We determined the fractional uncertainty in age at fixed mass by perturbing the input physics using the Yale Rotating Evolution Code. We modeled uncertainties in assumed helium abundance, heavy-element mixture, nuclear reaction rates, equation of state, opacity, model atmosphere, and heavy element diffusion rate as in \citet{vanSaders2012}. We added these uncertainties in quadrature to find the total fractional age uncertainty due to the choice of input physics for a RGB star to be $<\pm5\%$. This value was converted to a $<\pm1.5\%$ uncertainty in mass by defining a mass-age-composition relationship for first-ascent RGB stars using a grid of Dartmouth stellar evolution tracks \citep{Dotter2008}. Because of the rapid evolution along the RGB, the age difference between $\log g=1.0$ and $\log g=3.0$ changes the mass by $<0.3\%$ at fixed composition. This yielded a range of expected stellar masses as a function of metallicity, which may be directly compared to SR-based masses. Our total theoretical $1\sigma$ errors for halo and thick-disk stars are $\pm0.020$ M$_\odot$ and $\pm0.035$ M$_\odot$, respectively. The mean value of the expected mass varies slowly with metallicity at fixed age: a $+0.3$ dex shift in [Fe/H] at fixed T$_\mathrm{eff}$ increases the expected mass by only $\Delta M\lesssim 0.02$ M$_\odot$.

\section{Results and Discussion}\label{sec:Results}

\begin{figure}
\plotone{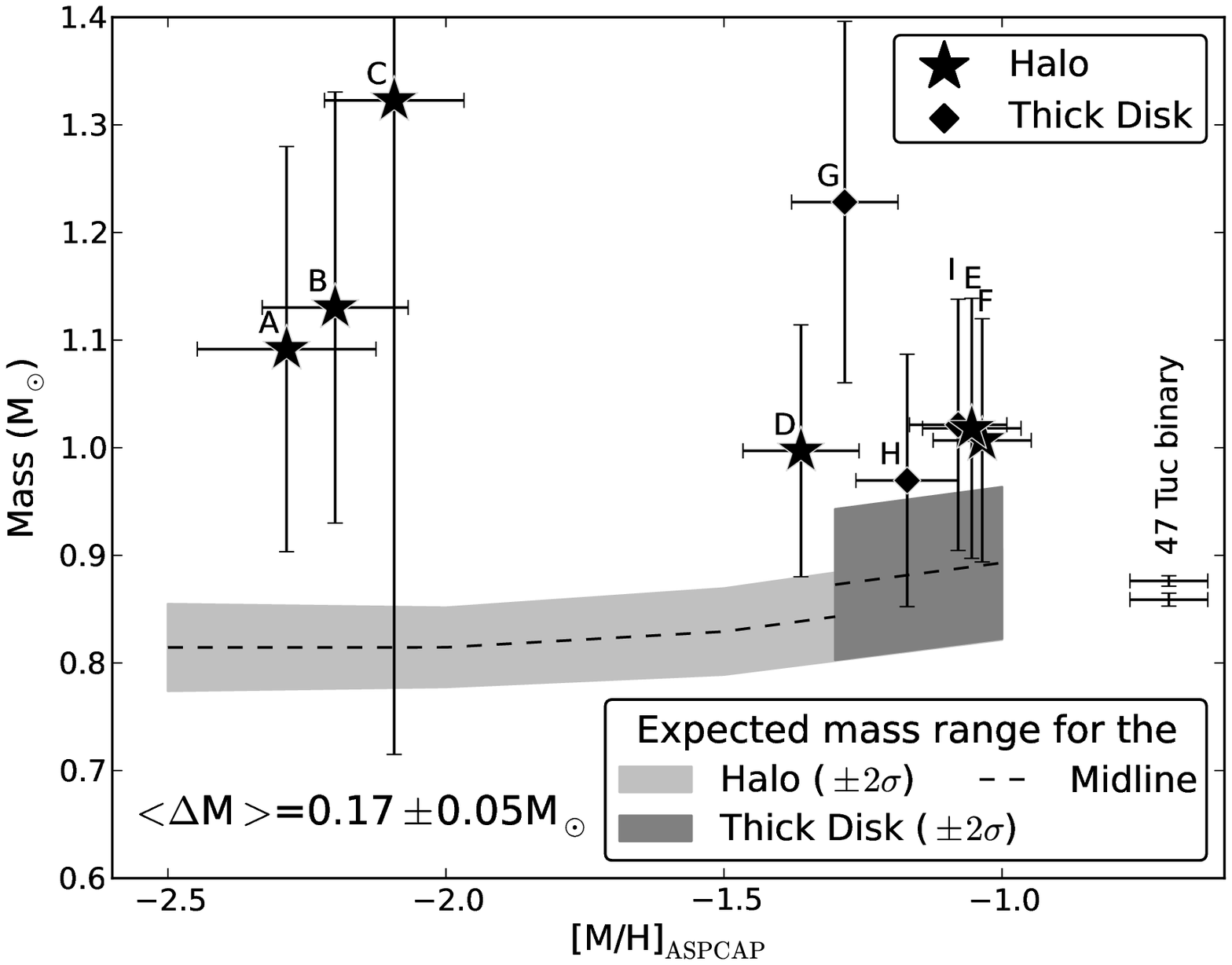}
\caption{Thick-disk (diamonds) and halo (stars) SR masses, calculated using $\Delta\nu$ and $\nu_\mathrm{max}$ from the APOKASC Catalog, versus metallicity. Compare the range of theoretically allowed masses for the halo and thick disk (light and dark gray bands, respectively) with dynamical masses from a metal-poor binary (see \S\ref{sec:Expectations and Models}).}\label{fig:mass-metallicity_Catalog}
\end{figure}

We computed SR masses and uncertainties using Equation \ref{eq:mass scaling relation}, taking as inputs spectroscopic temperature, $\Delta\nu$, and $\nu_\mathrm{max}$ and their uncertainties from the APOKASC Catalog. We compare these SR masses with our expectations for the halo and thick disk (\S\ref{sec:Expectations and Models}) in Figure \ref{fig:mass-metallicity_Catalog}. All of the SR-based mass estimates lie above the expected range. We computed $\left<\Delta \mathrm{M}\right>$, the weighted mean of the difference between the SR mass and the midline of the corresponding halo/thick-disk theoretical band at the appropriate metallicity, to be $0.17\pm0.05$ M$_\odot$ (standard error of the mean). We conservatively compare stars with an ambiguous kinematic classification to the wider thick-disk band. Weights include uncertainties in the SR masses and 1$\sigma$ uncertainties in the theoretical expectations. This $>3\sigma$ mass difference is significant and could result from some combination of (1) a problem with the physics in the stellar models; (2) problems in the definition or measurement of $\Delta\nu$, $\nu_\mathrm{max}$, or T$_\mathrm{eff}$; (3) a stellar populations effect; or (4) a breakdown in the underlying SR when extrapolated from the Sun to other stars. Point (1) was addressed in \S\ref{sec:Expectations and Models}. We examine the remaining possibilities below.

\subsection{SR Inputs}\label{sec:Measurement}

\begin{figure}
\plottwo{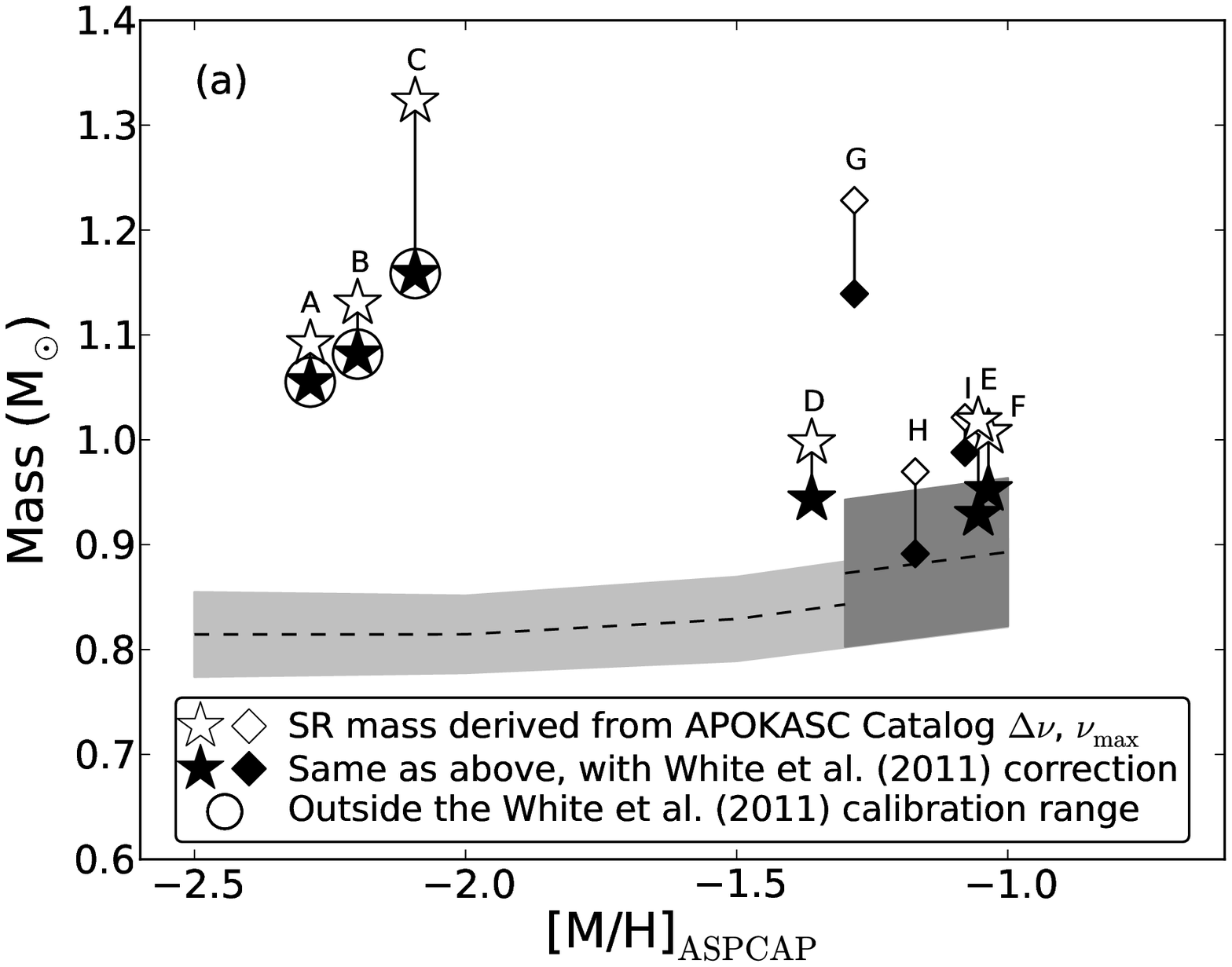}{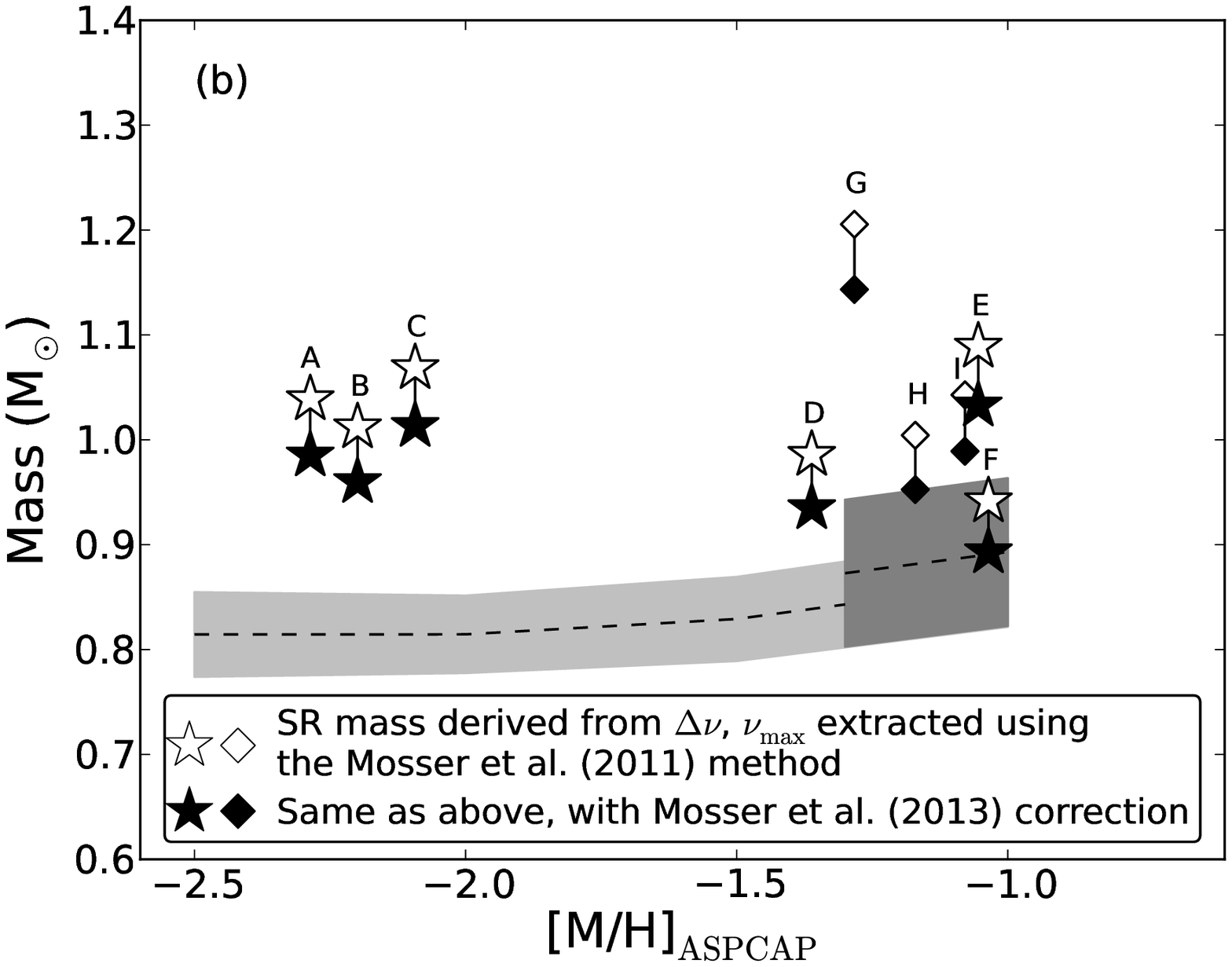}
\caption{SR mass versus metallicity; expected mass ranges are from Figure \ref{fig:mass-metallicity_Catalog}. \textit{Left:} Comparison of SR masses calculated using $\Delta\nu$ and $\nu_\mathrm{max}$ from the APOKASC Catalog with (black) and without (white) the \citet{White2011_Temperature_Correction} correction applied. Circles indicate where the \citeauthor{White2011_Temperature_Correction} correction has been extrapolated beyond its calibration range. \textit{Right:} Comparison of SR masses calculated using $\Delta\nu$ and $\nu_\mathrm{max}$ from the COR method (white) with the asymptotic correction applied (black).}\label{fig:mass-metallicity_corrections}
\end{figure}

Systematic effects in the SR inputs (T$_\mathrm{eff}$, $\Delta \nu$, and $\nu_\mathrm{max}$) could impact mass estimates. To explain the entire offset as a shift in the temperature scale requires an 11\% decrease in the spectroscopic temperature. The best-fit synthetic spectra show good agreement with the observed wings of the hydrogen lines at the adopted temperatures; this fit is degraded by adopting $\sim500$ K cooler temperatures. Additionally, we derive photometric temperatures using the infrared flux method (IRFM), which is more closely related to the fundamental definition of T$_\mathrm{eff}$. Because reddening is the largest uncertainty, we compute IRFM temperatures \citep{GonzalezHernandez2009} under three different reddening assumptions (Table \ref{table:Results}). The adopted spectroscopic temperature scale falls at the cool end of the range and is barely consistent with zero reddening in the Kepler field. The combined evidence therefore points toward a correction that would systematically increase the temperature and mass estimates.

Systematics in seismic parameters come in two flavors: differences between various methods of extracting $\Delta\nu$ and $\nu_\mathrm{max}$ from the power spectrum and theoretically motivated corrections.
The APOKASC catalog combines a variety of asteroseismic analysis methods and provides a representative uncertainty that characterizes the differences between the measured $\Delta\nu$ and $\nu_\mathrm{max}$. The large uncertainties for KIC8017159, for example, reflect the difficulties of measuring average values for luminous giants where only a few modes are available \citep{Mosser2013_Luminous_RGB}. Using different methods (e.g., \citealt{Kallinger2010}) to extract $\Delta\nu$ and $\nu_\mathrm{max}$ can shift $\left<\Delta \mathrm{M}\right>$ by as much as 0.04 M$_\odot$.

There are two published theoretically motivated corrections that impact Equation \ref{eq:delta nu scaling relation}, described in \citet[][]{White2011_Temperature_Correction} and \citet[][]{Mosser2013}. \citet{White2011_Temperature_Correction} found a temperature-dependent correction to Equation 1 for stars near solar metallicity; we combine this with Equation 2 to find an effective correction to the SR mass. We note that three stars are either more luminous or cooler, and all are more metal-poor, than those modeled by \citet{White2011_Temperature_Correction}. Figure \ref{fig:mass-metallicity_corrections}a shows that the \citet{White2011_Temperature_Correction} correction shifts stars to lower masses ($\sim6$\%), assuming the spectroscopic temperature scale. With this correction, $\left<\Delta \mathrm{M}\right>=0.10\pm0.04$ M$_\odot$. However, we caution that the effect depends on the absolute temperature scale. A hotter temperature scale (motivated by photometry) could reduce the magnitude or even reverse the sign of the change. Metallicity and evolutionary state could also impact the temperature correction \citep[see][]{Miglio2013_Differential_Population_Studies}.

We do not apply the \citet{Mosser2013} asymptotic correction directly to the Catalog $\Delta\nu$ because the correction is currently only calibrated for use with the COR method. Using the COR method to derive $\Delta\nu$ and $\nu_\mathrm{max}$ and computing SR masses with its associated solar reference values ($\Delta \nu_\odot=135.5$, $\nu_{\mathrm{max},\odot}=3104$ $\mu$Hz) yields $\left<\Delta \mathrm{M}\right>=0.16\pm0.05$ M$_\odot$, where the weights are set by the Catalog uncertainties. Applying the asymptotic correction reduces the SR masses by $4(\zeta-\zeta_\odot) = 5\%$, where $\zeta$ is the asymptotic correction factor. This calibration lowers $\left<\Delta \mathrm{M}\right>$ to $0.11\pm 0.05$ M$_\odot$ (Figure \ref{fig:mass-metallicity_corrections}b). We avoid combining this asymptotic correction with the \citet{White2011_Temperature_Correction} correction because they are highly correlated. For the considered set of stars, both the \citet{White2011_Temperature_Correction} and \citet{Mosser2013} corrections to the SR improve agreement with expectations.

\subsection{Stellar Populations}\label{sec:Stellar Populations}

Stars with a different history could masquerade as halo stars. Late accretion of in-falling satellite galaxies or stellar mergers (e.g., blue stragglers)
could be progenitors of a population of low-metallicity stars that are younger or more massive than their chemistry or kinematics suggest.  Blue stragglers compose only a few percent of the sample \citep{Andronov2006}.
We examined the \textit{Kepler} light curves for merger-induced rotational modulation and the APOGEE spectrum for rotational broadening, but found no detectable signatures.

KIC5858947 and KIC7265189 have luminosities below the horizontal branch and are asteroseismically confirmed as RGB stars from their mixed modes pattern \citep[e.g.,][]{Stello2013}. The remainder have ambiguous evolutionary state classifications. If some of our stars are helium core or shell burning, their current mass would be less than the RGB precursor by $\Delta\mathrm{M}\sim 0.15$ M$_\odot$ \citep{Lee1990}, a mass difference larger than the mass range predicted for halo RGB stars. Contamination of our sample by non-RGB stars would therefore strengthen our results.

\subsection{A Correction to the $\nu_\mathrm{max}$ Scaling}\label{sec:Frequency of Maximum Power}

\begin{figure}
\plotone{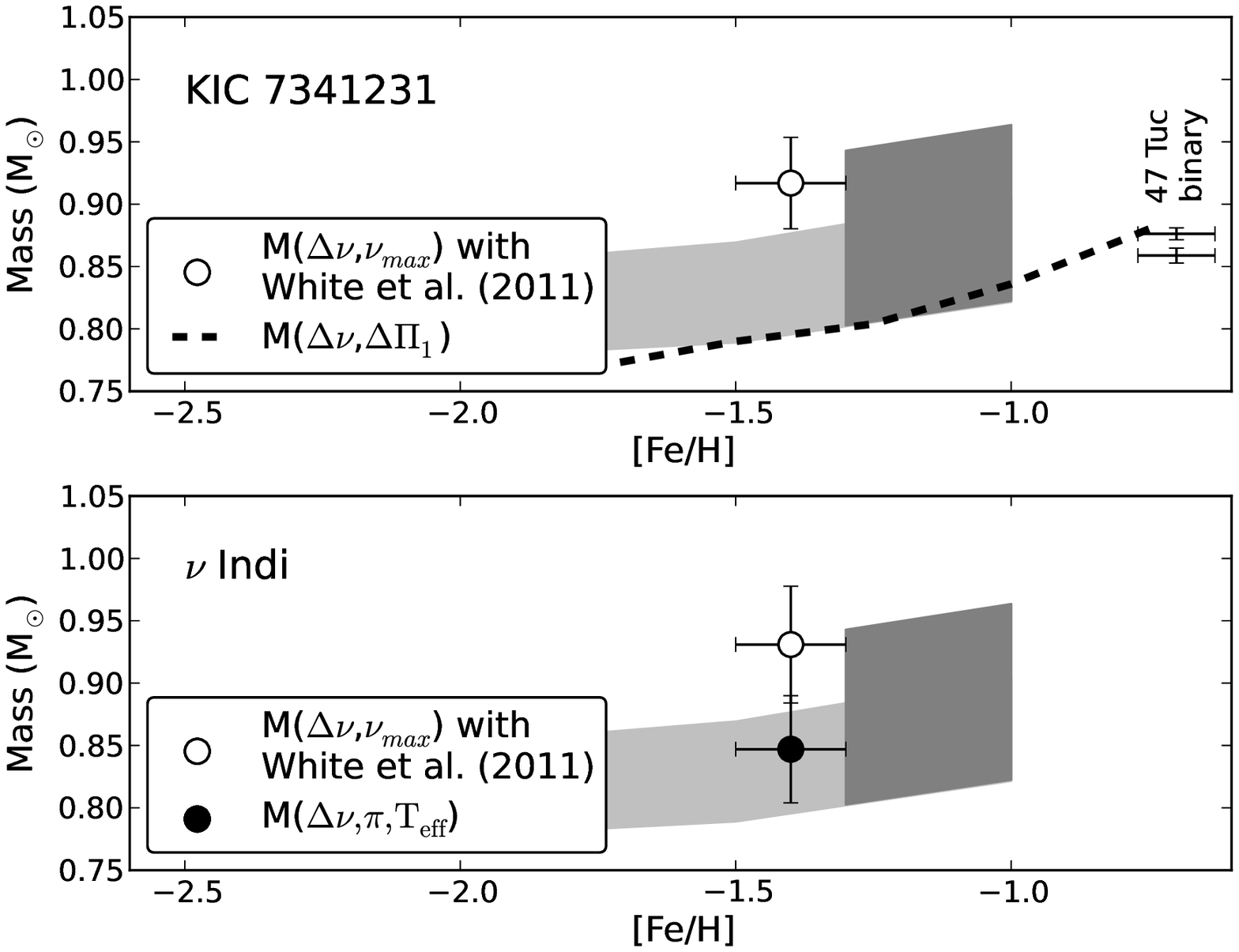}
\caption{Comparison of mass from SRs (with \citet{White2011_Temperature_Correction} corrections; white) and other techniques (black) for metal-poor stars from prior studies. Because KIC7341231's metallicity is not well-constrained, \citet{Deheuvels2012}\ provide a locus of possible best-fit masses. Expected mass ranges are from Figure \ref{fig:mass-metallicity_Catalog}.}\label{fig:literature seis}
\end{figure}

A 16\% decrease in mass would bring the uncorrected SR masses into agreement with the halo mass expectations. From Equation \ref{eq:mass scaling relation}, this could be achieved by a 4\% correction to the effective $\Delta\nu$, a 6\% correction to $\nu_\mathrm{max}$, or a combination of both. Equation \ref{eq:delta nu scaling relation} has a theoretical foundation: the asymptotic value of $\Delta\nu$ is related to the sound travel time across a star and therefore is tied to the mean stellar density. \citet{White2011_Temperature_Correction} and \citet{Mosser2013} have proposed modifications to the $\Delta\nu$ SR that affect mass on the 5\% level. \citet{Belkacem2011} argued for a physical interpretation of the $\nu_\mathrm{max}$ SR using the Mach number. Having $\nu_\mathrm{max}$ scale with the acoustic cutoff frequency (Equation \ref{eq:nu max scaling relation}) is a reasonable empirical approximation at least for solar-metallicity stars. However, it would not be surprising to find deviation from the empirical Equation \ref{eq:nu max scaling relation} when far from solar conditions. Metallicity could influence $\nu_\mathrm{max}$ through mode excitation and damping and opacity-driven changes in convective properties. To progress, we need to separately test the two SRs.

We estimate the sensitivity of mass estimates to $\nu_\mathrm{max}$ by comparing different analysis techniques. Previous studies have performed asteroseismic analyses of two metal-poor low-luminosity giants: $\nu$ Indi \citep{Bedding2006} and KIC7341231 \citep{Deheuvels2012}. We compute SR masses from the reported $\Delta\nu$ and $\nu_\mathrm{max}$, including the \citeauthor{White2011_Temperature_Correction} correction. Both studies also computed mass using additional constraints, independent of $\nu_\mathrm{max}$. \citeauthor{Bedding2006}\ constrained the luminosity of $\nu$ Indi with \textit{Hipparcos} parallax measurements. \citeauthor{Deheuvels2012} performed detailed modeling of individual frequencies in KIC7341231. Figure \ref{fig:literature seis} compares the SR masses with the masses derived by these other techniques. Interestingly, these SR masses are systematically higher ($0.08\pm0.06$ M$_\odot$ for $\nu$ Indi and $0.12\pm0.04$ M$_\odot$ for KIC7341231) than the masses derived from $\nu_\mathrm{max}$-independent fitting techniques. This evidence suggests that Equation \ref{eq:nu max scaling relation} may also require correction terms.

\section{Conclusions}\label{sec:Conclusions}

We identified six halo and three metal-poor thick-disk giants in the \textit{Kepler} field. Using independent constraints on the mass of halo and thick-disk stars, we performed the first test of asteroseismic SR masses in the metal-poor regime. We find that SR masses calculated with APOKASC Catalog parameters are $\left<\Delta \mathrm{M}\right>=0.17\pm0.05$ M$_\odot$ higher than expected for metal-poor stars. Published modifications of the $\Delta\nu$ SR reduce inferred masses by as much as 5\%. Additionally, masses derived for RGB stars from $\nu_\mathrm{max}$-independent methods are systematically lower than those from SR. This motivates future detailed frequency analyses of APOKASC metal-poor stars. Additionally, theoretical models from \citet{White2011_Temperature_Correction} suggest a metallicity-dependence in Equation \ref{eq:delta nu scaling relation} for RGB stars over the range [Fe/H$]=-0.2$ to $+0.2$. These theoretical predictions should be extended to [Fe/H$]<-1$ and lower $T_\mathrm{eff}$. Similarly, the reliability of the $\nu_\mathrm{max}$ determination and the impact of the $\nu_\mathrm{max}$-scaling on mass estimates requires investigation. We will use a larger sample of halo stars from additional APOGEE observations to better understand this mass offset.

\acknowledgements
Funding for SDSS-III has been provided by the Alfred P. Sloan Foundation, the Participating Institutions, the National Science Foundation, and the U.S. Department of Energy Office of Science. The SDSS-III web site is http://www.sdss3.org/.

SDSS-III is managed by the Astrophysical Research Consortium for the Participating Institutions of the SDSS-III Collaboration including the University of Arizona, the Brazilian Participation Group, Brookhaven National Laboratory, Carnegie Mellon University, University of Florida, the French Participation Group, the German Participation Group, Harvard University, the Instituto de Astrofisica de Canarias, the Michigan State/Notre Dame/JINA Participation Group, Johns Hopkins University, Lawrence Berkeley National Laboratory, Max Planck Institute for Astrophysics, Max Planck Institute for Extraterrestrial Physics, New Mexico State University, New York University, Ohio State University, Pennsylvania State University, University of Portsmouth, Princeton University, the Spanish Participation Group, University of Tokyo, University of Utah, Vanderbilt University, University of Virginia, University of Washington, and Yale University. 

We thank Andrea Miglio and the APOKASC team for helpful discussions. C.R.E., J.A.J., and M.P.\ acknowledge support by AST-1211673, T.C.B.\ by PHY 08-22648: Physics Frontier Center/Joint Institute for Nuclear Astrophysics (JINA), and S.M.\ by the NASA grant NNX12AE17G. 
S.H.\ acknowledges support from the Netherlands Organization for Scientific Research and ERC Starting Grant \#338251 Stellar Ages. 
A.S.\ is partially supported by the MICINN grant AYA2011-24704. Funding for the Stellar Astrophysics Centre is provided by The Danish National Research Foundation (Grant agreement No. DNRF106). V.S.A.\ is supported by the ASTERISK project (ASTERoseismic Investigations with SONG and {\it Kepler}) funded by the European Research Council (Grant agreement No. 267864).

\bibliographystyle{apj}
\bibliography{Epstein2014_Metal_Poor.bbl}

\newgeometry{margin=0.6cm} 
\begin{landscape}
\begin{deluxetable}{llcccccccccccccrrrrr}
\tabletypesize{\tiny}
\setlength{\tabcolsep}{0.02in} 
\tablewidth{0pc}
\tablecaption{Stellar Properties for the Metal-Poor APOKASC Sample}
\tablehead{\colhead{} & \colhead{} & \colhead{} & \colhead{} & \multicolumn{3}{c}{GHB2009 IRFM T$_\mathrm{eff}$}  & \colhead{Typical} & \colhead{} & \colhead{} & \colhead{} & \colhead{} & \colhead{} & \colhead{} & \colhead{} & \colhead{} & \colhead{} & \multicolumn{3}{c}{Asteroseismic Scaling Relations,} \\
 \colhead{} & \colhead{} & \colhead{\citeauthor{Schlegel1998}} & \colhead{} & \multicolumn{3}{c}{\underline{\ \ \ Assuming E(B-V)=\ \ \ }} & \colhead{IRFM T$_\mathrm{eff}$} & \multicolumn{3}{c}{\underline{APOGEE Spectroscopic Parameters}}  &  \colhead{} & \multicolumn{3}{c}{\underline{\ \ \ \ \ \ \ \ \ \ Kinematics\ \ \ \ \ \ \ \ \ \ }} &\multicolumn{2}{c}{\underline{Measured Seismic Parameters}} & \multicolumn{3}{c}{\underline{\ \ \ Assuming Spectroscopic T$_\mathrm{eff}$\ \ \ }}  \\
\colhead{} & \colhead{KIC} & \colhead{(\citeyear{Schlegel1998})} & \colhead{KIC} & \colhead{Zero} & \colhead{Schlegel} & \colhead{KIC} & \colhead{$\sigma_\mathrm{rand}$} & \colhead{T$_\mathrm{eff}$} & \colhead{$\log g$} & \colhead{[M/H]} & \colhead{Distance} & \colhead{U} & \colhead{V} & \colhead{W} & \colhead{$\Delta \nu$} & \colhead{$\nu_{max}$} & \colhead{Mass} & \colhead{Radius} & \colhead{$\log g$} \\
\colhead{} & \colhead{ID} & \colhead{E(B-V)} & \colhead{E(B-V)} & \colhead{(K)} & \colhead{(K)} & \colhead{(K)} & \colhead{(K)} & \colhead{(K)} & \colhead{(dex)} & \colhead{(dex)} & \colhead{(kpc)} & \colhead{(km/s)} & \colhead{(km/s)} & \colhead{(km/s)} & \colhead{($\mu$Hz)} & \colhead{($\mu$Hz)} &  \colhead{(M$_\odot$)}  &  \colhead{(R$_\odot$)} & \colhead{(dex)} }
\startdata
\multicolumn{10}{l}{\textbf{Halo}}\\
A	&	7191496	&	0.11	&	0.12	&	4930	&	5133	&	5167	&	100	&	4899	$\pm$	176	&	2.01	$\pm$	0.39	&	-2.29	$\pm$	0.16	&	2.45	$\pm$	0.24	&	-67	$\pm$	16	&	-260	$\pm$	8	&	-72	$\pm$	23	&	2.46	$\pm$	0.06	&	16.83	$\pm$	0.72	&	1.09	$\pm$	0.19	&	14.87	$\pm$	1.01	&	2.13	$\pm$	0.05	\\
B	&	12017985	&	0.1	&	0.1	&	4989	&	5178	&	5187	&	110	&	4862	$\pm$	163	&	1.99	$\pm$	0.33	&	-2.20	$\pm$	0.13	&	1.1	$\pm$	0.11	&	145	$\pm$	15	&	-180	$\pm$	1	&	-82	$\pm$	8	&	2.62	$\pm$	0.08	&	18.59	$\pm$	0.71	&	1.13	$\pm$	0.20	&	14.43	$\pm$	1.08	&	2.17	$\pm$	0.04	\\
C	&	8017159	&	0.07	&	0.12	&	4649	&	4764	&	4847	&	100	&	4586	$\pm$	162	&	1.10	$\pm$	0.29	&	-2.09	$\pm$	0.13	&	3.3	$\pm$	0.78	&	-280	$\pm$	46	&	-288	$\pm$	9	&	-26	$\pm$	24	&	0.64	$\pm$	0.07	&	3.08	$\pm$	0.13	&	1.32	$\pm$	0.61	&	38.90	$\pm$	8.70	&	1.38	$\pm$	0.05	\\
D	&	11563791	&	0.08	&	0.11	&	4892	&	5041	&	5095	&	90	&	4820	$\pm$	131	&	2.30	$\pm$	0.20	&	-1.36	$\pm$	0.10	&	0.99	$\pm$	0.07	&	13	$\pm$	7	&	-277	$\pm$	3	&	81	$\pm$	10	&	4.95	$\pm$	0.11	&	41.82	$\pm$	0.81	&	1.00	$\pm$	0.12	&	9.05	$\pm$	0.46	&	2.52	$\pm$	0.03	\\
E	&	11181828	&	0.07	&	0.14	&	4771	&	4884	&	5014	&	100	&	4702	$\pm$	121	&	2.16	$\pm$	0.24	&	-1.05	$\pm$	0.09	&	2.23	$\pm$	0.17	&	-70	$\pm$	81	&	-90	$\pm$	9	&	177	$\pm$	37	&	4.13	$\pm$	0.09	&	33.49	$\pm$	0.72	&	1.02	$\pm$	0.12	&	10.29	$\pm$	0.53	&	2.42	$\pm$	0.03	\\
F	&	5858947	&	0.11	&	0.11	&	5051	&	5257	&	5247	&	120	&	4820	$\pm$	119	&	2.73	$\pm$	0.20	&	-1.04	$\pm$	0.09	&	0.7	$\pm$	0.05	&	147	$\pm$	10	&	-160	$\pm$	4	&	14	$\pm$	9	&	14.41	$\pm$	0.26	&	174.41	$\pm$	4.18	&	1.01	$\pm$	0.11	&	4.45	$\pm$	0.21	&	3.14	$\pm$	0.03	\\
\multicolumn{10}{l}{\textbf{Thick Disk}}\\
G	&	7019157	&	0.06	&	0.16	&	4936	&	5042	&	5219	&	80	&	4754	$\pm$	129	&	2.02	$\pm$	0.20	&	-1.28	$\pm$	0.10	&	2.33	$\pm$	0.19	&	91	$\pm$	15	&	-37	$\pm$	5	&	-18	$\pm$	13	&	3.43	$\pm$	0.08	&	27.68	$\pm$	0.80	&	1.23	$\pm$	0.17	&	12.39	$\pm$	0.71	&	2.34	$\pm$	0.03	\\
H	&	4345370	&	0.12	&	0.16	&	4617	&	4811	&	4872	&	100	&	4726	$\pm$	125	&	2.25	$\pm$	0.20	&	-1.17	$\pm$	0.09	&	1.32	$\pm$	0.1	&	-19	$\pm$	13	&	-94	$\pm$	8	&	-71	$\pm$	20	&	4.03	$\pm$	0.09	&	31.81	$\pm$	0.69	&	0.97	$\pm$	0.12	&	10.29	$\pm$	0.54	&	2.40	$\pm$	0.03	\\
I	&	7265189	&	0.07	&	0.11	&	4915	&	5031	&	5107	&	120	&	4903	$\pm$	138	&	2.72	$\pm$	0.23	&	-1.08	$\pm$	0.09	&	1.24	$\pm$	0.09	&	93	$\pm$	7	&	-19	$\pm$	4	&	113	$\pm$	10	&	8.43	$\pm$	0.17	&	85.01	$\pm$	1.78	&	1.02	$\pm$	0.12	&	6.40	$\pm$	0.31	&	2.83	$\pm$	0.03	\\\hline
\enddata\label{table:Results}
 \tablecomments{Pixel data (Q0-Q12) are available for KIC4345370, KIC5858947, and KIC8017159; otherwise public data (Q0-Q8) were used.}
\end{deluxetable}
\end{landscape}
\restoregeometry

\end{document}